\documentclass[11pt]{article}

\usepackage{latexsym, amscd, amsfonts, eucal, mathrsfs, amsmath, amssymb, amsthm, xr,  makeidx, stmaryrd}
\usepackage{cancel}
\usepackage{thmtools,thm-restate}
\usepackage{hyperref}   
\usepackage{epigraph}
\usepackage{graphicx}
\usepackage[natbibapa]{apacite}
\usepackage[usenames]{color}
\usepackage{verbatim}
\usepackage{tensor}
\usepackage[shortlabels]{enumitem}
\usepackage{tikz-cd}
\usetikzlibrary{decorations.pathmorphing, patterns}
\usepackage{subcaption}
\usetikzlibrary{backgrounds}
\usepackage{float}
\usepackage{multirow}
\usepackage{ragged2e}
\usepackage{setspace}
\usepackage{pgfplots}
\usepackage{titling}

\date{}

\begin{document}

\setlength{\droptitle}{-5.5cm}

\title{Galileo's Ship and the Relativity Principle \\\small{Forthcoming in \textit{Noûs}}}

\author{Sebastián Murgueitio Ramírez \\{\small Purdue University}}

\vspace{-14cm}

\maketitle
\vspace{-1cm}

\begin{abstract}


It is widely acknowledged that the Galilean Relativity Principle, according to which the laws of classical systems are the same in all inertial frames in relative motion, has played an important role in the development of modern physics. It is also commonly believed that this principle holds the key to answering why, for example, we do not notice the orbital velocity of the Earth as we go about our day. And yet, I argue in this paper that the precise content of this principle is ambiguous: standard presentations fail to distinguish between two principles that are ultimately inequivalent, the "External Galilean Relativity Principle" (EGRP) and the "Internal Galilean Relativity Principle" (IGRP). I demonstrate that EGRP and IGRP play distinct roles in physics and that many classical systems that satisfy IGRP fail to satisfy EGRP.  I further show that the Relativity Principle introduced by Einstein in 1905---which is not restricted to classical systems---also leads to two inequivalent principles. I conclude by noting that the phenomenon originally captured by Galileo's famous ship passage is much more general than contemporary discussions in the philosophy of symmetries suggest. 




\end{abstract}
\vspace{0.6cm}
\noindent
\textbf{}


\tableofcontents

\section{Introduction}

Consider ripples in a pond, a pendulum clock, the sounds produced by the vibrations of a guitar string, a dart flying towards a dartboard, or any other phenomenon on Earth whose laws are well-described by Newtonian mechanics (at least at a certain level of approximation).  The behavior of any of these systems does not depend on whether the date is, say, July 4th, 2026 or January 4th, 2027, even though the Earth's orbital speed will be about 2200 miles per hour faster on the second date compared to the first one. For example, even if Alberta is an excellent musician, they would not be able to detect a difference in the sound of the guitar on these two dates. But why? Why is it that changes in the orbital velocity of the Earth (and the guitar) produce no changes in the sound created by the guitar? 

A natural answer is that the waves associated with the guitar’s strings are a classical phenomenon and thus satisfy the so-called ``Galilean Relativity Principle" according to which, roughly for now, \textit{the laws of classical systems are the same in any inertial frame}, where an inertial frame is one in which a body subject to no forces moves with constant speed in a straight line (i.e., a frame in which Newton's first law holds). This principle is supposed to apply to all of the phenomena mentioned above, whether it be the ripples in a pond, the clock, or the vibrations in the guitar's strings. Applied to the case at hand, the principle implies that the laws for classical waves (such as string waves and acoustic waves) remain the same when switching from the (approximately) inertial frame associated with Alberta's living room on July 4th, 2026, to the (approximately) inertial frame associated with their living room on January 4th, 2027 (where the second frame is moving at around 2200 miles per hour relative to the first frame). Since these laws are the same in both frames, it is no surprise that the vibrations of the strings, and hence the sounds produced by them, are the same too. Indeed, this is similar to Einstein's remarks about an analogous case: 

\begin{quote}
    We should expect [if the Galilean Relativity Principle is not true], for instance, that the note emitted by an organ-pipe placed with its axis parallel to the direction of travel [of the Earth] would be different from that emitted if the axis of the pipe were placed perpendicular to this direction \cite[pp. 25-26]{einsteinRelativitySpecialGeneral2015a}.
\end{quote}

So far so good. But there is a problem! The laws of classical waves are \textit{not} invariant under \textit{Galilean boosts}, the transformations linking different inertial frames in relative motion in Newtonian mechanics.\footnote{In \S \ref{Systems that do not satisfy EGRP}, I will explain that classical waves are similar to electromagnetic waves in that they are preserved by Lorentz boosts, the boosts used in Special Relativity (e.g., the kind of boost that makes a rod contract relative to an observer who is not moving with the rod).} This is a problem because it seems to follow from this lack of invariance under Galilean boosts that the laws for classical waves do not satisfy the Galilean Relativity Principle. In particular, it seems to follow that the laws of these kinds of systems are distinct in different inertial frames in relative motion. Thus, the answer to our question, as well as Einstein's explanation of the organ-pipe case, needs to be revisited. In fact, the original question seems more puzzling now: if the laws governing classical waves are not the same in inertial frames in relative motion, why is it that changes in the orbital velocity of the Earth (and the guitar) produce no changes in the sound created by the guitar?

In this paper, I will argue that there is a hitherto unnoticed ambiguity in standard presentations of the Galilean Relativity Principle (GRP). In particular, I argue that standard presentations fail to distinguish between two principles that are ultimately inequivalent, the ``External Galilean Relativity Principle" (EGRP) and the ``Internal Galilean Relativity Principle" (IGRP).  I will argue that these two principles, both widely endorsed in physics practice, are associated with two different ways in which the laws of classical systems can be the same in inertial frames in relative motion. The distinction between EGRP and IGRP will help us see that, despite what is typically suggested in the literature, there is a sense according to which a system can satisfy the Galilean Relativity Principle even if it lacks laws that are invariant under  Galilean boosts---that is, even if these boosts are not \textit{symmetries} of its laws (I will explain these concepts in due course).  Acoustic and string waves are good examples, but we will see that there are many others.

Clarifying what the GRP says is important not only for philosophical purposes (e.g., for thinking about what laws and symmetries are), but also for historical and scientific reasons. GRP played a significant role in the development of modern physics, from Galileo's famous ship thought experiment to Newton's Corollary V in the \textit{Principia} to Einstein's development of special relativity (see \S\ref{ShipsTrains}). Arguably, to better understand the development of Newtonian Mechanics as well as Special and General Relativity, it is important to better understand GRP, which is central to all of these theories. Also, as mentioned above, the principle is taken by many to explain a rather basic fact about our universe, namely, that systems inside bigger systems in approximate uniform motion behave the same way regardless of the velocity of the bigger system. We encounter this fact when traveling in trains or airplanes or simply when standing in the street while our planet moves around the Sun at several thousand miles per hour. But this kind of explanation does not work if the principle is understood as EGRP, as will become clear later. 

The structure of the paper is as follows: First, in \S\ref{ShipsTrains}, I start with a brief historical overview of GRP, beginning with Galileo's famous ship passage and ending with the Relativity Principle (RP) introduced by Einstein in his 1905 paper on special relativity (RP is a more general version of GRP, a version that is not restricted to classical systems). Then, in \S\ref{Internal vs External Relativity}, I show that standard presentations of GRP are ambiguous between two different principles, both of which fall under the label ``Galilean Relativity Principle" (here I also show that there is an analogous ambiguity affecting RP). In \S \ref{Systems that do not satisfy EGRP}, I prove that these two principles---EGRP and IGRP---are not equivalent, by showing that some classical systems compatible with IGRP are incompatible with EGRP. In particular, I show in this section that classical systems that have laws that are not invariant under Galilean boosts satisfy IGRP but not EGRP. In this section, I also clarify how the acoustic Doppler effect relates to the distinction between IGRP and EGRP. Finally, in \S\ref{Beyond the classical world}, I show that the phenomenon originally captured by Galileo’s famous ship passage is much more general than is normally assumed, and I briefly indicate why this generality creates some problems for the widely defended view that there is a strong link between the symmetries of a system and the observations performed on that system. 

\section{From Ships to Trains}
\label{ShipsTrains}

As explained in the introduction, a standard way of framing the Galilean Relativity Principle is the following: the laws of classical systems are the same in all inertial frames in relative motion.\footnote{For instance, the Stanford Encyclopedia of Philosophy entry on Symmetries defines the principle like this (my emphasis): ``the laws of physics are invariant under Galilean boosts" \citep{brading_temporal_2023}. On the other hand, the Stanford Encyclopedia of Philosophy entry titled ``Space and Time: Inertial Frames" defines the principle in this slightly different way (my emphasis): ``\textit{mechanical experiments} will have the same results in a system in uniform motion that they have in a system at rest" \citep{disallerobertSpaceTimeInertial2020}. It will be clear in \S\ref{Internal vs External Relativity} that these two definitions can be read as tracking to some extent the external and the internal reading of the Galilean Relativity Principle, respectively.}  I will ultimately argue that there is an ambiguity in this and similar presentations of the principle, but for now, I will leave this complication aside and focus on some key moments in the historical development of the principle. From now on, to keep the language simpler, \textit{when I talk about more than one inertial frame, I assume that they are in relative motion along a certain direction} (of course,  one could also talk about two inertial frames that are shifted or rotated, but those are not the cases that will concern me here). For example, think of the (approximately) inertial frame associated with a train parked in the station and the inertial frame associated with the train when moving at 100 miles per hour in a straight line. Hence, using this terminology, we can define the principle in this way:

\begin{quote}
   \textbf{Galilean Relativity Principle (GRP)}: the laws of classical systems are the same in all inertial frames. 
\end{quote}

This principle is named after Galileo because, on Day Two of his \textit{Dialogues}, Galileo presented (through Salviati) an important thought experiment that illustrates a principle very much like this one. The thought experiment, which was used by Galileo to illustrate that we cannot tell that the Earth is moving simply by observing how objects behave around us, goes as follows (my emphasis):

\begin{quote}
    Shut yourself up with some friend in the main cabin below decks on some large ship, and have with you there some flies, butterflies, and other small flying animals. Have a large bowl of water with some fish in it; hang up a bottle that empties drop by drop into a wide vessel beneath it. With the ship standing still, observe carefully how the little animals fly with equal speed to all sides of the cabin. The fish swim indifferently in all directions; the drops fall into the vessel beneath; [...] When you have observed all these things carefully [\dots], have the ship proceed with \textit{any speed you like, so long as the motion is uniform and not fluctuating this way and that}. \textit{You will discover not the least change in all the effects named, nor could you tell from any of them whether the ship was moving or standing still.} In throwing something to your companion, you will need no more force to get it to him whether he is in the direction of the bow or the stern, with yourself situated opposite. [\dots] the butterflies and flies will continue their flights indifferently toward every side, nor will it ever happen that they are concentrated toward the stern, as if tired out from keeping up with the course of the ship, from which they will have been separated during long intervals by keeping themselves in the air \citep[p. 186]{galileiDialogueConcerningTwo1967}.
\end{quote}

Although our current concept of a classical system and an inertial frame were developed after Galileo's time (for one, Galileo's inertial motion concerned circular, not rectilinear motion [e.g., see \citet[pp. 31-32]{galileiDialogueConcerningTwo1967} and \citet{chalmersGalileanRelativityGalileo1993}]), it is easy to see why so many have believed that this passage expressed something very similar to the idea that the laws of classical systems are the same in all inertial frames.\footnote{Although this is a rather common belief, we will soon see that there are reasons to think that Galileo was not intending to restrict his principle to classical or mechanical systems.}  After all, we can associate an (approximately) inertial frame with the ship at rest in the port, and a different inertial frame (with a different velocity) with the ship when sailing \textit{uniformly}. In both frames, all the objects in the cabin seem to behave in the same manner, and, in particular, they all seem to satisfy the same laws. In Galileo's words, ``You will discover not the least change in all the effects named, nor could you tell from any of them whether the ship was moving or standing still."

Some decades later, a version of the same idea appeared in Newton's work, now presented as a corollary to the laws of motion of the new physics developed in the \textit{Principia}.\footnote{See \citet[\S1.3]{disallerobertSpaceTimeInertial2020} for more details about the early history of the principle, including Christiaan Huygens's own version from 1656.} The corollary in question reads as follows:

\begin{quote}
    Corollary V: When bodies are enclosed in a given space, their motions in relation to one another are the same whether the space is at rest or whether it is moving uniformly straight forward without circular motion \citep[p. 423]{newtonPrincipiaMathematicalPrinciples1999}. 
\end{quote}

Note that by ``space,'' Newton means a system that is approximately closed, for example, the cabin of a ship. Indeed, in the derivation of the corollary that he presents just afterward, Newton gives the example of a ship: ``This is proved clearly by experience: on a ship, all the motions are the same with respect to one another whether the ship is at rest or is moving'' \citet[p. 423]{newtonPrincipiaMathematicalPrinciples1999}. Note also that the motion relative to the shore of the bodies inside a ship is different if the ship is moving or if it is anchored, so it is important when stating this principle to distinguish the motion of the bodies with respect to other bodies within the same space (e.g., the ship) from their motion with respect to bodies outside that space (e.g., the shore), as Newton explicitly does here (he is talking of ``their motions in relation to one another" in the \textit{same space}).\footnote{\citet[Ch. 4]{brownPhysicalRelativitySpacetime2005} argues that this distinction between the motion of bodies relative to other bodies inside a space, and their motion relative to bodies external to that space, gives empirical meaning to GRP.} Of course, Galileo knew this as well, which is why his example concerns the motion of objects inside the cabin below deck. 

From our modern perspective, it seems natural to read this corollary in the \textit{Principia} as stating that the laws of Newtonian systems (not just their \textit{motions}) are the same regardless of what inertial frame is adopted (where, once again, we have associated an inertial frame to a ``space" such as the ship when anchored).\footnote{\label{BarbourBrown} Newton’s derivation of this corollary appeals to both the second law and an example involving objects colliding with one another, but it would be an understatement to say that the derivation is lacking. Indeed, as both \citet[pp. 31-32]{barbourAbsoluteRelativeMotion1989} and \citet[p. 37]{brownPhysicalRelativitySpacetime2005} have pointed out, Newton's derivation omits the crucial premise according to which the forces acting on all the bodies ``in a given space" (e.g., inside the ship) do not depend on the velocity of such space.} That is, from our modern perspective it is natural to read this corollary as presenting GRP or at least an earlier version of it. 

It would take over two centuries after the publication of the first edition of the \textit{Principia} before someone coined the term ``principle of relativity." It is believed the term was first used by Henri Poincaré in  an address delivered before the International Congress of Arts and Science in St. Louis, on September 24th, 1904 \citep[pp. 207-208]{browneGalileiProposedPrinciple2020}.\footnote{Poincaré's address was reprinted some months later in \textit{The Monist}. It was titled ``Principles of Mathematical Physics" \citeyearpar{poincarePRINCIPLESMATHEMATICALPHYSICS1905}. See \href{https://www.gutenberg.org/files/38267/38267-h/38267-h.htm}{https://www.gutenberg.org/files/38267/38267-h/38267-h.htm} for the original program of the congress, which included well-renowned scholars in many disciplines such as Ludwig Boltzmann, Gaston Darboux, Wilhelm Ostwald, and Thomas Woodrow Wilson, who would become President of the United States nine years later.} In the address, Poincaré lists six important physical principles of the time, among which he includes the conservation of energy and the following one (my emphasis):

\begin{quote}
    The principle of relativity, according to which the \textit{laws} of physical phenomena should be the same, whether for an observer fixed, or for an observer carried along in a uniform movement of translation; so that we have not and could not have any means of discerning whether or not we are carried along in such a motion \citep[p. 5]{poincarePRINCIPLESMATHEMATICALPHYSICS1905}
\end{quote}

Notice that in contrast to Newton's corollary V, this principle is stated explicitly in terms of the laws of physical systems and not their motions, making it more similar to modern formulations. It is also worth stressing that Poincaré does not restrict the principle to mechanical or classical systems. For instance, in the same piece, Poincaré considers how electrodynamics seems to threaten the principle \citeyearpar[pp. 9-10]{poincarePRINCIPLESMATHEMATICALPHYSICS1905}, discusses how the optical experiments performed by Michelson and Morley offered an experimental confirmation of it \citeyearpar[pp. 10-11]{poincarePRINCIPLESMATHEMATICALPHYSICS1905}, and  concludes that ``the principle of relativity has been valiantly defended in these latter times, but the very energy of the defense proves how serious was the attack" \citeyearpar[pp. 11]{poincarePRINCIPLESMATHEMATICALPHYSICS1905}.

Even though Poincaré first used the term ``relativity principle," it was Einstein who brought the principle to prominence a few months later in his famous paper ``On the Electrodynamics of Moving Bodies," published on June 30, 1905. In that paper, Einstein says (my emphasis):

\begin{quote}
    the same \textit{laws} of electrodynamics and optics will be valid for all frames of reference for which the equations of mechanics hold good [i.e., for all inertial frames]. We will raise this conjecture (the purport of which will hereafter be called the ``Principle of Relativity'') to the status of a postulate \citep{einsteinElectrodynamicsMovingBodies1905}.
\end{quote}
 
In section 2 of the same paper, Einstein defines the principle in this way (my emphasis): 

\begin{quote}
    The \textit{laws} by which the states of physical systems undergo change \textit{are not affected}, whether these changes of state be referred to the one or the other of two systems of co-ordinates in \textit{uniform translatory motion}. \citep[\S2]{einsteinElectrodynamicsMovingBodies1905}
\end{quote}

Notice that Einstein, like Poincaré, presents this principle in terms of the laws of physical systems rather than their motions. Also, Einstein is explicit that the principle should encompass ``electrodynamics and optics" in addition to ``mechanics," and uses the language of frames of reference (in the first passage) or coordinate systems in uniform motion (in the second one), which is standard in modern presentations. This is not a coincidence, of course, as modern presentations typically follow Einstein's approach. 

Einstein's influential popular book on relativity, published in 1920, also has an illustrative presentation of the principle:

\begin{quote}
    If, relative to K [an inertial system], K' is a uniformly moving co-ordinate system devoid of rotation, then natural phenomena run their course with respect to K' according to exactly the same general laws as with respect to K. This statement is called the principle of relativity (in the restricted sense).\footnote{Einstein uses ``restricted sense"  because by 1920, when this book was published, he had already developed the theory of general relativity, and in doing so, had introduced a more general principle of relativity that included accelerating frames. In his book, Einstein called it the ``General Principle of Relativity" \citeyearpar[\S 22]{einsteinRelativitySpecialGeneral2015a}. For a recent and careful philosophical discussion of it, see \citet{Lehmkuhl2022-LEHTEP}.} As long as one was convinced that all natural phenomena were capable of representation with the help of classical mechanics, there was no need to doubt the validity of this principle of relativity \citep[p. 24]{einsteinRelativitySpecialGeneral2015a}.
\end{quote}

For ease of reference, we can summarize the relativity principle Einstein is introducing in these three passages in this manner:

\begin{quote}
    \textbf{Relativity Principle (RP)}: the laws of physical
systems are the same in all inertial frames.
\end{quote}

Notice that GRP is a special case of RP, as the former says ``classical systems" and the latter employs the more generic term ``physical systems." 

At this point, it is worth flagging that the term ``Galilean Relativity Principle" can be misleading in at least one important way (other issues with the term will be noted later). The fact that the term is defined as being restricted to classical or mechanical systems makes it sound as if Galileo himself had intended to limit the ship passage to these kinds of systems. But this seems false for at least three reasons. First, notice that in the ship passage, Galileo does not discuss any standard mechanical system at all, such as a lever, a pulley, or an inclined plane. Instead, he illustrates his principle by appealing to ``biological systems" such as butterflies, fish, and a person throwing an object (he also uses smoke and water drops). Second, as \citet[p. 36]{brownPhysicalRelativitySpacetime2005} notes, if Galileo thought that non-mechanical systems such as light rays or magnets were not compatible with the ship thought experiment, he probably would have said so at some point (Galileo discusses both light and magnets in various parts of the \textit{Dialogues}).\footnote{A similar point can be made about Newton's corollary V \citep[p. 36]{brownPhysicalRelativitySpacetime2005}.} And third, in the introduction to the book, Galileo says that (my emphasis) ``\textit{all} experiments practicable upon the earth are insufficient measures for proving its mobility, since they are indifferently adaptable to an earth in motion or at rest" \citep[p. 6]{galileiDialogueConcerningTwo1967}. For these reasons, it seems more accurate to read Galileo's ship passage as presenting a principle closer in scope to RP than to GRP. Having said this, to avoid unnecessary confusion, I will still follow standard terminology and keep using ``GRP" for the version of RP that is restricted to classical systems.

Finally, consider a more recent presentation of the Relativity Principle found in the first volume of the Feynman lectures. There, when talking of Newton's corollary V, Feynman makes the following comment (my emphasis):

\begin{quote}
    This principle is that \textit{the laws of physics will look the same} whether we are standing still or moving with a uniform speed in a straight line. For example, a child bouncing a ball in an airplane finds that the ball bounces the same as though he were bouncing it on the ground.  [\dots] This is the so-called \emph{Relativity Principle}. As we use it here [in the context of Newtonian mechanics] we shall call it ``Galilean relativity” to distinguish it from the more careful analysis made by Einstein, which we shall study later \citep[ch. 10]{feynmanFeynmanLecturesPhysics1963}.
\end{quote}

Curiously, despite what he says in the last sentence, when Feynman goes on to discuss the Relativity Principle in the context of the theory of Special Relativity in chapter 15 of the same volume, he appeals to Newton's corollary V again and never considers Einstein's own definition!

\section{Internal vs External Relativity}
\label{Internal vs External Relativity}

\subsection{Two senses of having the same laws}
\label{Two senses of having the same laws}

Recall that Feynman characterizes GRP by saying that ``the laws of physics will look the same whether we are standing still or moving with a uniform speed in a straight line." But what does it mean to say that the laws \textit{look} the same, exactly? And what does it mean to say, as in the definitions of GRP and RP, that the laws \textit{are the same} in all inertial frames? Perhaps Einstein's formulation can help here. In particular, recall that he characterizes RP as saying that ``natural phenomena run their course with respect to $K'$ according to exactly the same general laws as they do with respect to $K$," where $K$ and $K'$ refer to two different inertial frames. Presumably, this means that if we study the behavior of a physical system $S$ using frame $K$, we find it to obey the same laws as the ones it does obey in frame $K'$. Still,  this is not completely clear. Are we supposed to consider a single system as studied from the perspective of two different frames in relative motion with respect to one another, or are we supposed to consider two copies or instances of the very same type of system, where one copy is co-moving with frame $K$ and the other one is co-moving with frame $K'$?  To better understand this question and see why it matters, consider a very simple example.

Say that Sara, while standing inside a train, throws a dart at a board hanging on the front wall of the cabin (so she is throwing the dart in the same direction as the motion of the train). In the cabin, there is a camera that records the motion of the objects and is also capable of making measurements of speed, time, and acceleration. Imagine, furthermore, that the cabin of the train is made out of glass, so observers outside can see what is happening in the cabin. A second camera is also installed outside on a post at the train station. When the cabin is passing by the post with a constant speed of 100 kph, Sara throws the dart. We then collect the video clips from the two cameras and compare them. As expected, both recordings indicate the same vertical acceleration (i.e., $g$), no horizontal acceleration, and the same time for the dart's flight. They also indicate different measurements for the traveled distance and the horizontal velocity of the dart, as the external camera sees the dart as moving with the velocity of the train in addition to the velocity relative to the cabin (the internal camera only tracks the latter). Crucially, both sets of measurements are consistent with Newton's second law. In particular, they both measure null horizontal acceleration, which is expected from this law if there is no horizontal force. And they both measure the same vertical acceleration, which is, again, expected given that the dart's weight is (approximately) constant as it falls.   Hence, even though the recorded positions for any time will be different because the velocities are different, the laws obeyed by the system are the same (to be more precise, these positions will be seen to satisfy the very same differential equations). This example illustrates one way in which the same phenomena (the dart's behavior) ``run their course with respect to $K'$ according to exactly the same general laws as with respect to $K$.''

Now, imagine Sara throwing a dart inside a train's cabin, but this time, we will consider two different throws, and we will forget about the external camera. First, when the train is parked at the station, Sara will throw the dart, and the camera in the cabin will record the motion. Then, when the train is moving at 100 kph, Sara will throw the dart again, with the same initial speed with respect to the cabin, and at the same angle and from the same height. We then collect the video clips from the two throws and compare them.  Just like in the case discussed above, both clips indicate the same vertical acceleration, no horizontal acceleration, the same time for the flight, and, more generally, the recorded positions are seen to trace a trajectory that satisfies the differential equations associated with Newton's second law. Hence, we say that both throws of the dart, each moving at a different speed with respect to the ground, obey the same laws. This is then a second type of instance in which the same phenomena ``run their course with respect to $K'$ according to exactly the same general laws as with respect to $K$." Importantly, in contrast to the previous case,  the two recordings show the same distance, the same horizontal velocity, and, in general, the very same \textit{motion} as a function of time.  In fact, in this second case, we could not tell one video clip apart from the other one!  

So there seem to be at least two different ways of understanding a phrase such as ``obeys the same laws in different inertial frames'' and thus two different readings of the Galilean Relativity Principle (and, for the same reason, two readings of the Relativity Principle, but we will focus on GRP for the time being).  In one reading, which I call ``external,'' the phrase means obeying the same laws for a given \textit{single} system simultaneously studied from two (or more) inertial frames. In practice, this means that one can adequately capture the kind of behavior of the system with the very same equations in both frames (e.g., one can use the equations for projectile motion in two frames).  In a second reading, which I call ``internal,'' the phrase in question means that if we study a system from the perspective of a certain inertial frame and we boost both the frame and the system together (e.g., we perform a Galilean boost on the train and the dart), then the laws of the boosted system from the perspective of the boosted frame are the same as the laws of the original system from the perspective of the original frame (notice that this assumes that the initial conditions are the same; we will come back to this shortly). In this second case, not only is the \textit{type} of behavior the same (e.g., parabolic motion), but the specific behaviors are also the same (e.g., one recovers the exact same parabola).  

It is important to note that when considering the internal sense, we don't necessarily need to think of a single frame and a single system that are being boosted together, like a single cabin and a single dart. Instead, we can consider two instances of the very same type of system, such as two identical darts. For example, let's say that at 3:00 pm, Sara throws a dart with initial conditions $C$ while standing in a train that is moving in a straight line at 100 kph. At the same time (or different time, it does not matter!), Carlos throws an identical dart with the same initial conditions $C$, but he is standing inside a train moving at 5 kph in a straight line. According to the internal interpretation of ``obeys the same laws in different inertial frames,'' both Carlos and Sara will observe that their darts behave according to the same laws (i.e., they will observe their positions as a function of time to satisfy the same differential equations). Moreover, they will see that the darts have the exact same motion (i.e., their positions as a function of time seem exactly alike). So the internal reading ought to be interpreted more broadly than just applying to a single system that is studied first in one frame and then boosted to a new frame. If there are already two instances of the same system moving at different velocities, there is no need to boost anything according to the internal reading.  

Although the distinction between the internal and the external interpretations of the phrase in question might seem clear enough, offering a slightly more precise formulation would be useful. To do so, we need to say a bit more about the initial conditions of a system. Think of the initial conditions $C$ for a system in an inertial frame $F$ as corresponding to whatever the (relevant) properties of that system are, as measured with respect to an object (e.g., a device or an observer) at rest in the origin of $F$ at some ``initial" time $t$ (what we take the initial time to be is, of course, arbitrary).\footnote{One can easily generalize the discussion by considering a device or observer in motion in $F$, or one not in the origin.} For example, the initial conditions for the dart in the frame associated with the train include things such as the initial height, initial velocity, and initial angle at which the dart is thrown, where these properties are specified with respect to some object in the cabin, say a camera attached to the cabin's floor (we take that camera to be located at the origin of the frame associated with the cabin when the latter is moving uniformly). Recall that from the perspective of another device, such as the camera at rest in the train station, the very same dart, when thrown in a moving train, has a different initial velocity precisely because the train is moving with respect to this external camera. Hence, for cases that fall under the external interpretation, it will generally be the case that the initial conditions for the very same system will be different depending on the frame under consideration precisely because of the relative motion of the frames (this is why we do not require the same initial conditions in the external interpretation). However, in cases concerning the internal interpretation, the initial conditions as measured by different devices at rest in different frames will be the same (at least in the cases of interest). For convenience, I will use ``$C_{K}$" to represent the initial conditions as specified with respect to the origin of frame $K$. For example,  $C_{K}$ will represent the initial velocity, initial height, and initial angle of the dart when measured with respect to a camera, which is at the origin of $K$ and remains at rest in $K$. 

We are now in a better position to see how the two interpretations of a phrase such as ``obeys the same laws in different inertial frames'' correspond to two different relativity principles for mechanics. 

\begin{quote}
     \textbf{External Galilean Relativity Principle (EGRP)}: Consider an arbitrary classical system $S_1$ with initial conditions $C_{K}$ in the inertial frame $K$. $S_1$ behaves according to the same laws in $K$ and in $K'$, where $K'$ is an inertial frame in uniform motion with respect to $K$ (in general, the initial conditions $C_{K'}$ in $K'$ will be different from those in $K$). To phrase it more precisely, the \textit{kind} of behavior of $S_1$ is captured by the same equations in $K$ and in $K'$. The \textit{specific} behavior will be, in general, different in $K$ and $K'$ (i.e., the trajectories will be different but they will still satisfy the same equations). 
\end{quote}

\begin{quote}
   \textbf{Internal Galilean Relativity Principle (IGRP)}: Consider an arbitrary classical  system $S_1$ of type $T$ with initial conditions $C_{K}$ in the inertial frame $K$. Take a second classical system  $S_2$ of type $T$ (it could be the very same $S_1$, now boosted, or it could be a duplicate)  with initial conditions $C_{K'}$, where $K'$ is an inertial frame moving uniformly with respect to $K$.  If the initial conditions are the same in the two frames (if $C_{K}=C_{K'}$), then the \textit{specific} behavior of $S_1$ in $K$ is the same as the \textit{specific} behavior of $S_2$ in $K'$. It trivially follows from this that the \textit{kind} of behavior of $S_1$ in $K$ is captured by the same equations as the ones that capture the \textit{kind} of behavior of $S_2$ in $K'$.
\end{quote}

To go back to our example, according to EGRP, the external camera will see the dart in Sara's cabin follow a different parabola compared to the one seen by the internal camera, so the specific behavior of the dart is different depending on the frame. However, the two cameras agree in that the two trajectories satisfy the differential equations for parabolic motion, and so agree about the kind of behavior of the dart. According to IGRP, if the initial conditions are the same, the dart in Sara's cabin will follow the exact same parabola as the dart in Carlos' cabin. Hence, the specific behavior of the dart is the same regardless of the frame, and so, trivially, the kind of behavior (parabolic motion) is also the same. A more succinct yet slightly technical way of describing the situation goes like this: in the case of EGRP, different inertial frames pick out different solutions for the same laws of a given system, whereas in the case of IGRP, different inertial frames pick out the very same solution for the same laws. 

 I would like to make three remarks about these principles. First, an important consequence of IGRP is that it would be impossible to tell if we are in an inertial frame $K$ or in another inertial frame $K'$ simply by looking at how a classical system behaves in our frame. For example, we can't tell that it is January as opposed to June simply by seeing how a pendulum sitting on my desk oscillates during these two months, even though one would associate (approximately) different inertial frames with my desk in these two months. Second, it is easy to see that the two interpretations of the phrase ``obeys the same laws in different inertial frames'' also lead to two different versions of the Relativity Principle (RP), an internal one (``IRP") and an external one (``ERP"). IRP (ERP) would be formulated just like IGRP (EGRP) except for the fact that it would use ``physical systems" instead of ``classical systems."
 
 Third, and finally, it is worth clarifying how the distinction between EGRP and IGRP relates to the distinction between active and passive transformations.  Say we study the trajectory of a single dart from the perspective of two inertial frames in the context of Newtonian mechanics, the frame of the external camera (at rest in the station) and the frame of the internal camera (moving at 100 kph in a straight line). If we know the trajectory $r(t)$ of the system in one frame as a function of time, we can derive the trajectory $r'(t)$ in the other one simply by performing a \textit{passive} (or coordinate) transformation that maps $r(t)$ to the new trajectory $r'(t)=r(t)-vt$, where $v$ represents the velocity between the two frames (in other words, this means that the coordinates in one frame are related to the ones in the other frame via a \textit{Galilean boost}). For example, if $r(t)$ corresponds to the parabola the internal camera records,  then $r'(t)=r(t)-vt$ would correspond to the parabola recorded by the external camera (the latter is stretched compared to $r(t)$, but it is still a parabola). The fact that both frames see the dart as following parabolic motion indicates that this passive transformation does not change the laws of the dart (as EGRP requires), or in slightly more technical jargon, it indicates that Galilean boosts are symmetries of the dart's laws. Now study the dart when thrown in the cabin at rest, and then when thrown in the cabin that is moving at 100 kph.  In this case, one can go from the trajectory $r(t)$ seen in one frame to the trajectory $r''(t)$ seen in the other one by performing both (a) a passive transformation that maps $r(t)$ to  $r'(t)=r(t)-vt$  \textit{and} (b) an \textit{active} transformation that maps $r'(t)$  to $r''(t)=r'(t)+vt$. The passive one encodes information about how the coordinates in one frame relate to the coordinates in the other one (just as it was in the case of EGRP), whereas the active one represents the change in the velocity of the dart (the dart in the cabin at rest is moving with a different velocity than the dart in the cabin moving at 100 kph). In the end, these two transformations exactly cancel out, and so the \textit{net} transformation relating the trajectory of the dart in one frame to the trajectory in the other one ends up being trivial;  $r(t)$ is mapped to itself. In other words, if one increases the velocity of the observer and the velocity of the dart by the same amount, the relative motion between the dart and the observer remains the same. Thus,  the two frames see identical parabolas for the dart, $r(t)=r''(t)$, just as IGRP requires. In short, according to EGRP in the context of Newtonian mechanics, trajectories are related by a passive transformation of the form $r(t) \mapsto r(t)-vt$  whereas according to IGRP they are identical, $r(t) \mapsto r(t)$ (``$a \mapsto b$" means that $a$ is being mapped to $b$).
 
For completeness, let me note that if instead of Newtonian mechanics we had considered Special Relativity, the transformations relating trajectories in one frame to trajectories in another frame would have been different. In particular, if we were considering EGRP in a relativistic context, then instead of assuming Galilean boosts of the form $r(t) \mapsto r(t)-vt$, we would have to assume Lorentz boosts of the form $r(t) \mapsto \gamma(r(t)-vt)$, where  $\gamma = \frac{1}{\sqrt{1 - v^2/c^2}}$ (at low velocities relative to the speed of light, $\gamma \approx 1$, and the two transformations coincide). In the case of IGRP, it would remain true that $r(t) \mapsto r(t)$ because the passive Lorentz boost would cancel out the active one. This serves to illustrate the often overlooked fact that the Galilean Relativity Principle, either in the form of IGRP or EGRP, does not require different inertial frames to be connected by Galilean boosts specifically, despite what the name might suggest.\footnote{\citet[p. 36]{brownPhysicalRelativitySpacetime2005} also stresses this, although without distinguishing between EGRP and IGRP.} In fact, notice that nowhere in the definition of EGRP and IGRP is it specified what kind of transformation connects different inertial frames.  Still, when evaluating if a given system satisfies these principles, it is important to specify if we are in a non-relativistic context in which inertial frames are related by Galilean boosts, or if we are in a relativistic one in which they are related by Lorentz boosts. After all, as we will see in the next sections, some systems might satisfy EGRP (or ERP) only in one of the two contexts. Unless noted otherwise, I will keep discussing EGRP and IGRP in non-relativistic contexts for now.

 \subsection{Is IGRP the obvious reading?}
 \label{Is IGRP the obvious reading?}

Let me end this section by addressing a potential concern.  One might think that it is obvious that the Galilean Relativity Principle is the one expressed by IGRP because IGRP directly captures the idea that things look the same ``from within a given closed space" regardless of the state of uniform motion of such space.  Indeed, a version of IGRP does seem to be the one alluded to in Galileo's passage, or in Newton in Corollary V, or in Einstein's passage about the organ-pipe discussed in the introduction, and also the one hinted at by Feynman in the passage discussed earlier. IGRP is also the principle that seems to be discussed by philosophers of physics such as \citet{belotGeometryMotion2000}, Brown \citeyear{brownPhysicalRelativitySpacetime2005}, Norton \citeyear{nortonWhyConstructiveRelativity2008} and \citet{disallerobertSpaceTimeInertial2020}. So why even suggest that EGRP is a plausible reading of this principle? 

It turns out that many physicists have defined or talked about the Galilean Relativity Principle in ways that more closely resemble EGRP than IGRP. For example,  \citet[p. 105]{emamCovariantPhysicsClassical2021} defines GPR in the following way: ``Any two observers moving at constant speed and direction with respect to one another will obtain the same results for all physical experiments.'' On its own, this might not seem like enough to distinguish IGRP from EGRP, but just a few lines later, the author goes on to add that the two observers will disagree about properties such as the velocity \citeyearpar[p. 106]{emamCovariantPhysicsClassical2021}.  Hence, Emam is thinking of EGRP, not IGRP. Or consider this passage by \citet[p. 55]{zeeFearfulSymmetrySearch2007}: ``relativistic invariance says that two observers in relative motion must arrive at the same physical laws, in spite of the fact that they differ in their measurements of various physical quantities [e.g., speed]."  Here, ``relativistic invariance" seems to be Zee's own term for GRP, as he introduces it just two lines after referencing Galileo's ship \citeyearpar[p. 54]{zeeFearfulSymmetrySearch2007} and in the same section where he discusses the impossibility of measuring absolute motion using mechanical experiments (this is also the section in which Zee discusses Einstein's own relativity principle). Hence, given the quoted definition of relativistic invariance, it seems that Zee is thinking of GRP in terms of EGRP and not IGRP. Similarly, in the chapter dedicated to the Relativity Principle (RP) of the very influential textbook on relativity by \citet{taylorSpacetimePhysics1992}, there is a section clarifying that even though different frames agree about the laws for a system, they disagree about certain properties of the system, such as its velocity or space separation \citeyearpar[\S3.2]{taylorSpacetimePhysics1992}. In the section after that one, the authors go on to emphasize that according to RP, the laws and physical constants are required to be the same in all inertial frames, but properties like the velocity can vary \citep[\S 3.3]{taylorSpacetimePhysics1992}. So Taylor and Wheeler are thinking about the Relativity Principle in the external sense captured by ERP.

Furthermore, even in Einstein's 1905 paper, presumably the most influential paper on the Relativity Principle, the external perspective is central. For example, according to Einstein, for RP to be compatible with the fact that the speed of light is independent of the state of motion of the source (the so-called ``light postulate"), it has to be the case that different observers in relative motion agree about the speed of any particular light ray (that is, the speed of a single light ray observed from two different frames).  Einstein goes on to prove this result by using the Lorentz transformations for a spherical wave and comes to the following conclusion (my emphasis):

\begin{quote}
    \textit{The} wave under consideration is therefore no less a spherical wave with velocity of propagation $c$ when viewed in the \textit{moving system}. This shows that our two fundamental principles [the Relativity Principle and the light-postulate] are compatible \citep[p. 8]{einsteinElectrodynamicsMovingBodies1905}.
\end{quote}

Hence, Einstein believes that for RP to be compatible with the light-postulate, it must be the case that the laws for a single light wave (he says ``\textit{the} wave") are the same in different inertial frames, which suggests that he is reading RP externally, in the way captured by ERP (here the relevant laws are that the speed of light is $c$ and that the wave is spherical). Indeed, notice that it would not be surprising at all that two different light beams, produced in two different train cabins in relative motion with respect to one another, are seen to have the same speed and to be spherical in their corresponding cabin. The surprising thing is that these properties of light are preserved when a single light beam is seen from two frames in relative motion! Similarly, relativistic effects such as length contraction or time dilation require adopting the external perspective. For instance, the contraction of a rapidly moving rod is only evident externally, from the perspective of a detector that is not moving together with the rod (this is why  \citet[\S 3.2]{taylorSpacetimePhysics1992} go on to say that RP does not require that different inertial frames agree about space separation or time intervals). 

Finally, we can offer a more general argument that EGRP constitutes a fairly widespread understanding of GRP.  It is very common to find discussions of GRP that present the principle, or even try to justify it, by showing that $F=ma$ remains invariant under the Galilean boost transformation $x \mapsto x-vt$ (e.g., see \citet[p. 9567]{clineInertialFramesReference2021}, \citet[ch. 15]{feynmanFeynmanLecturesPhysics1963}, \citet[ch. 3]{northPhysicsStructureReality2021}, and \citet[ch. 1]{susskindGeneralRelativityTheoretical2023}).\footnote{Crucially, these authors must assume that neither $m$ nor $F$ depends on the velocity, otherwise the equation is not invariant under boosts (see footnote \ref{BarbourBrown}). For the classical systems that we will study in this paper, this assumption holds.} But to think that GRP is the same as, or that it is justified by, the fact that $F=ma$ is invariant under $x \mapsto x-vt$ is to think about GRP along the lines of EGRP. To see this, recall from the end of \S\ref{Internal vs External Relativity} that in the context of Newtonian mechanics, EGRP involves a (passive) boost transformation of the form $x \mapsto x-vt$. This transformation tells us how $x$, the position variable in one inertial frame, is related to $x'=x-vt$, the position variable of another frame moving with velocity $v$ with respect to the first frame. So to show that $F=ma$ is invariant under $x \mapsto x-vt$ amounts to showing that the two frames agree about the fact that the system (one system) satisfies Newton's second law, even though they disagree about the position and velocity of the system at any given time. Thus, showing that $F=ma$ is invariant under $x \mapsto x-vt$ is a way of demonstrating that EGRP holds. Recall too that IGRP ultimately amounts to the trivial transformation $x \mapsto x$ because the passive transformation relating the coordinates in the frames cancels out the active one relating the two velocity states of the system (needless to say, $F=ma$ is invariant under $x \mapsto x$).  Hence, the fact that in their discussion of GRP none of these authors mention the trivial invariance of $F=ma$ under $x \mapsto x$ but rather focus on the more interesting invariance of $F=ma$ under $x \mapsto x-vt$ strongly suggests that they are understanding GRP along the lines of EGRP and not those of IGRP.\footnote{For exactly analogous reasons, the fact that scholars illustrate or explain RP by showing that non-trivial Lorentz boosts are symmetries of the wave equation for light in vacuum (as Einstein does in the passage discussed earlier) suggests that in relativistic contexts, ERP is widespread.}

\section{Systems that do not satisfy EGRP}
\label{Systems that do not satisfy EGRP}

Up until now, I have argued that there are two different readings of GRP, namely, EGRP and IGRP, and have shown that both seem to have ample support (there are also two readings of RP, ERP and IRP). The possibility remains, however, that EGRP and IGRP are ultimately equivalent principles. This section shows that this is not the case. The argument is simple: in non-relativistic contexts, many classical systems serve as counterexamples of EGRP but not of IGRP. In \S\ref{Beyond the classical world}, I also show that the same argument holds in relativistic contexts both for classical and non-classical systems.

\subsection{Guitars, springs, and planets}
\label{Guitar waves, springs and planets}

To begin with, consider again a system like the dart thrown inside a train's cabin, and let's focus on the horizontal motion for simplicity. As we noted earlier, even though the specific motion recorded by the external and internal cameras are different, both trajectories are consistent with Newton's second law for the case of projectile motion (no horizontal acceleration and constant vertical acceleration). As we also said, if we repeat the experiment at different uniform velocities of the cabin,  say when parked and when moving at 100 kph, we obtain in both cases a motion that looks exactly alike in the two cases, and, of course, a motion consistent with Newton's second law in the two cases. Hence, the dart is an example of a classical system that is consistent with both EGRP and IGRP.

For another simple example, consider a person pushing a chair inside the train cabin with force $F$, and suppose that there is no friction. Then, the external and internal cameras will show different specific motions for the chair (as the instantaneous velocity will be different), but the two motions will satisfy $F=m\frac{d^2}{dt^2}x$, Newton's second law for this particular system. Mathematically, this is reflected in the fact that $x \mapsto x-vt$ (a Galilean boost) is a symmetry of $F=m\frac{d^2}{dt^2}x$.\footnote{$F$ and $m$ are held constant.} It is also clear that if the person repeats the experiment at different uniform velocities of the cabin, say when it is parked and when it is moving in a straight line at 100 kph, they will obtain in both cases a motion for the chair that looks exactly alike (we must assume, of course, that the chair is pushed in the same way and also that its mass has not changed). So this seems to be another simple example of a case that seems to be consistent with both IGRP and with EGRP. 

It is tempting to infer from these two cases that we can generalize this pattern to any other classical system, but things are not so simple. Consider a slightly more complex system, namely, the vibrations of a string like those obtained when playing guitar. Imagine that Sara makes a string vibrate inside the cabin when the train is parked and then makes it vibrate when the train is moving at 100 kph by pulling the strings in the same way. For the camera inside, the vibrations of these two cases look exactly the same; the waves in the string will be seen as having the same amplitude and the same frequency. In particular, these waves will be seen to behave according to the very same solution of the so-called ``wave equation".\footnote{\label{classicalwave}The dynamical equation for waves in a string is 

 \begin{equation}
 \label{Wave}
    \frac{\partial^2 y}{\partial t^2} = \frac{T}{\mu} \frac{\partial^2 y}{\partial x^2},
\end{equation} where $T$ is the tension in the string, $\mu$ the string's density, $t$ time, $x$ the horizontal displacement and $y$ the vertical displacement. Here, $\sqrt{T/\mu}$ is the velocity of propagation of the wave in the string.}  So the vibrations in a string, and any other classical wave more generally (such as water waves, acoustic waves, seismic waves, etc), seem to be examples of classical systems consistent with IGRP. Indeed, this is precisely what Einstein seemed to be alluding to with the organ example presented in the introduction (also, since IGRP is a special case of IRP, all the systems that satisfy the former also satisfy the latter).  This is, of course, expected. We are familiar with the fact that regardless of the Earth's particular speed around the Sun, one can use the classical wave equation to model many kinds of classical waves! 

\begin{figure}
    \centering
    \includegraphics[width=1 \linewidth]{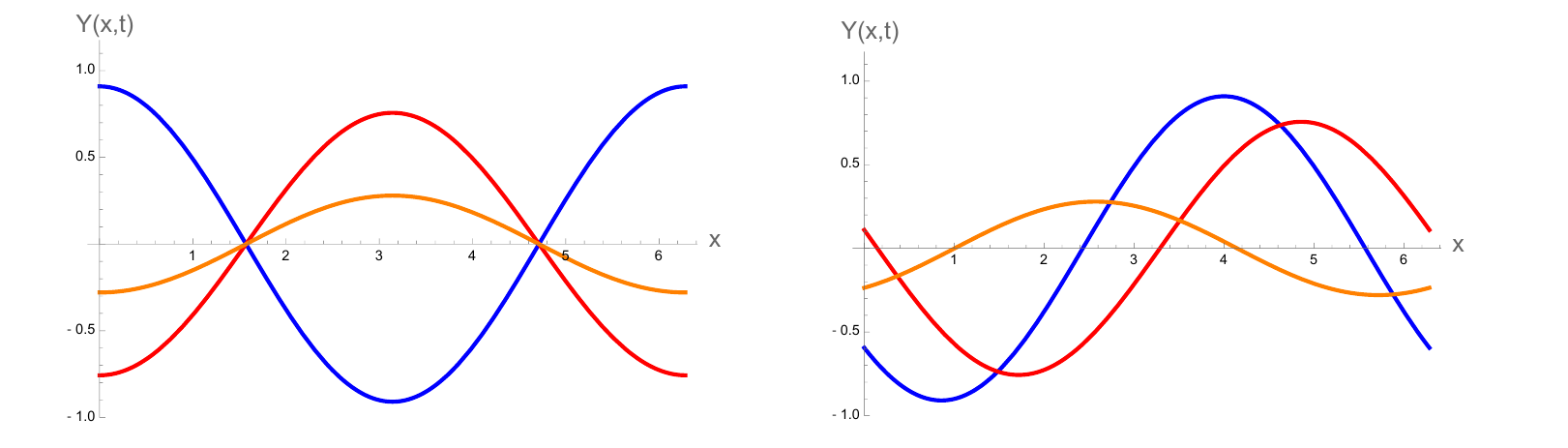}
    \caption{\small{On the left we have a stationary wave in the guitar as seen by the internal camera at three different times (the colors blue, red and orange correspond to the wave at three different times). On the right we have the shape that the external camera would see for the very same wave at the same time intervals. Notice that whereas the internal camera sees the wave going up and down (a signature of a stationary wave), the external camera sees it as moving both up and down and also moving along the $x$ axis. Crucially, the vibrations in the right figure do not even satisfy the wave equation (footnote \ref{classicalwave}) but a more complicated differential equation (footnote \ref{movingwave}). The particular functions plotted here are $y(x,t)=\sin (2t)\cos(x)$ for the left figure and $y(x,t)=\sin (2t)\cos(x-t)$ for the right one.}}
    \label{fig:waves}
\end{figure}

What about the behavior of the string's waves when seen from the perspective of the external camera as the train passes by? The external camera will see different \textit{specific} shapes or patterns for the vibrations precisely because of the velocity of the string with respect to the station. This is just as it was with the dart, whose specific motion looked different when seen from outside. However, the \textit{type} of behavior for the vibrations, as seen by this camera, is also different when compared with the one seen by the internal camera. In particular, the vibrations on the string as seen by the external camera will fail to satisfy the classical wave equation when the train is moving. Instead, these vibrations will be seen to satisfy a \textit{different} equation that has a term that captures the relative motion between the string and the frame (see figure \ref{fig:waves} for an illustration).\footnote{\label{movingwave}In particular, the vibrations will obey 
\begin{equation}
\label{TransformedWave}
\frac{\partial^2 y}{\partial t'^2} + 2v \frac{\partial^2 y}{\partial x' \partial t'} + \left(v^2 - \frac{T}{\mu}\right) \frac{\partial^2 y}{\partial x'^2} = 0,
\end{equation} where $v$ is the velocity of the string with respect to the external camera (not to be confused with the velocity of propagation of the wave in the string, which is given by $\sqrt{T/\mu}$). Notice that this differential equation is different from the one in footnote \ref{classicalwave}.}  Mathematically, this is reflected in the fact that $x \mapsto x-vt$ (a Galilean boost) is \textit{not} a symmetry of the classical wave equation. So, unlike the case of the dart, this is an instance in which the law of a classical system (as captured, e.g., by the differential equation used to model that system) is \textit{different} in the external and the internal frame. Hence, this is an instance of a system that is \textit{not} consistent with EGRP.

It turns out that many other classical systems that are consistent with IGRP are inconsistent with EGRP. For example, as examined in detail by \citet{murgueitioramirezSymmetriesSprings2024}, Hooke's law, the equation used to model ideal springs, is not invariant under Galilean boosts. This means that two observers looking at the very same spring inside the cabin, one from outside as the train passes by and one from the inside, will see it obey different laws, contrary to what EGRP says (the external observer will not see the spring as satisfying Hooke's law). Of course, as recorded by the internal observer, the spring will be seen to obey Hooke's law when the train is parked and also when it is moving uniformly, and so it is a classical system consistent with IGRP. For another example, note that Galilean boosts are not even symmetries of the equation for the Kepler problem that is used to describe the motion of a body in an inverse square force field, such as that of a planet moving around the Sun.\footnote{The equation is \begin{equation}
    \frac{d^2\mathbf{r}}{dt^2} + \frac{\mu}{r^3} \mathbf{r} = 0,
\end{equation} where $\mathbf{r}$ is the position vector and $\mu$ is a constant. This equation is \textit{not} invariant under boosts of the form $\mathbf{r} \mapsto \mathbf{r} -\mathbf{v}t$ (see \citet{princeLieSymmetriesClassical1981} for a discussion of the symmetries of this equation).} This might seem rather puzzling, as Newtonian gravitation is supposed to be a paradigmatic example of a theory whose laws are invariant under Galilean boosts. And the puzzle only increases if one notes that neither the gravitational force between the two bodies nor their acceleration depends on the inertial frame from which they are being observed. So, how come changes in the inertial frame, as when we go from the internal to the external camera,  give rise to changes in the equation used to model the gravitational behavior of the bodies? Indeed, we can raise a similar concern regarding ideal springs and classical waves. Neither the tension in the guitar's string nor the force on an object attached to an ideal spring depends on the velocity between the system and the inertial frame adopted to study them. So how is it that the differential equations used to model these systems depend on such velocity? 

The key to solving these puzzles relies on the fact that the equations used to model these and other classical systems that are inconsistent with EGRP in a non-relativistic context (i.e., systems whose laws are not invariant under Galilean boosts) do presuppose a special inertial frame. In the case of the spring, the equation assumes a frame that is at rest with respect to the equilibrium position. In the case of the vibrations in the string, a frame is adopted so that the string is initially at rest (so ``standing waves" are, in fact, \textit{stationary} in this frame). In the case of the Kepler problem, one assumes a frame in which the source of the potential is at rest. The same is true of the wave equation for sound propagation, which adopts a frame in which the medium (the air) is initially at rest, or the heat equation, which adopts a frame in which the material that is being heated, say a metallic rod, is not moving. In all these cases, the differential equations capture how the system behaves from the perspective of these particular frames, so it should not be that surprising that the equations are not invariant under Galilean boosts. In other words, given that scientists study many systems from the perspective of a lab frame in which the system (or a part of it) is initially at rest, it is not that surprising that the equations they end up using to model those systems only properly model the system in those particular frames; they do so by design.\footnote{This does not mean that scientists could not find the equations for other frames. They can easily do that by starting with the equations that assume the lab frame and then studying how they change under $r(t) \mapsto r(t)-vt$ (they will change, for Galilean boosts are not symmetries of those equations).} Note that this is not unlike the case of the wave equation for light, whose lack of invariance under Galilean boosts was interpreted by many scientists as suggesting that there was a privileged frame, namely, the one in which the ether was supposed to be at rest. Unlike the case of sound waves or waves in a string, however, scientists were not able to detect such a medium for light (see \citet{chengWhyNotSound2021} for a careful study comparing the case of sound waves to that of light in the context of GRP).

Notice that the lack of invariance under Galilean boosts of these laws is consistent with the fact that neither the accelerations nor the forces acting on the various objects depend on the velocity between them and the frame. If one were to compute the acceleration and the force of, say, a block attached to a spring as seen from a frame that is in uniform motion with respect to the equilibrium position, it would be the same as the acceleration and the force calculated from the rest frame of equilibrium \citep[\S6]{murgueitioramirezSymmetriesSprings2024}, and the same goes for the other systems such as those involving gravitational bodies. This illustrates the often overlooked fact that even if all inertial frames were to agree about the forces and accelerations for a given classical system, and so even if they were to agree about the fact that the system obeys $F=ma$, it might still happen that different inertial frames would disagree about the specific laws that the system satisfies, precisely because those laws might be sensitive to features of the system that do depend on its velocity with respect to the frame, such as whether the medium of the wave is initially at rest. This is why it is misleading to take simple cases like the dart or the chair when evaluating IGRP and EGRP, as these systems are rather special and satisfy both. 

Now, the reader might think that to say that the laws for these classical systems do presuppose a special inertial frame is problematic, for it seems to be at odds with the widely accepted belief that, in physics, any inertial frame is as good as any other for the description of nature---that is, no inertial frame is privileged. Indeed, this kind of answer seems problematic also on more philosophical grounds: the laws of nature are not supposed to be the kinds of things that are rather specific or that depend on arbitrary choices on the part of agents, such as a choice of frame for the observer or detector. For these reasons, one can object to the claim that the laws of some classical systems are not compatible with EGRP in this way: if it is true that the classical wave equation and many other equations that represent the laws of some classical systems do presuppose specific inertial frames, then, despite appearances, these equations do not capture the kind of regularities that achieve ``lawhood status." That is, these equations do not capture genuine laws. And if they do not capture genuine laws, then, the objection goes, they pose no challenge to EGRP. 

I think, however, that this kind of objection is lacking, as it is fueled by an ambiguity in the claim that no inertial frame is privileged. A frame in which the string is at rest is indeed ``privileged" from the point of view of classical waves, in the sense that it is the frame in which those waves behave in the way captured by the classical wave equation---for example, it is the frame in which standing waves are, indeed, stationary. But it is also true that no inertial frame is preferred over any other one in this more substantive sense: whatever the inertial frame in which the string is at rest is, waves in that frame do behave according to the very same equation. To use a previous example, no matter what the train's uniform velocity is (i.e., no matter the frame), as long as the string is pulled in the same way every time, the vibrations will look just the same inside the cabin every single time. This is precisely what we expect if IGRP is true, and presumably what many have in mind when saying that there is not a privileged inertial frame.

A similar thing can be said regarding the concern that the wave equation does not capture a genuine law. If it were true that the wave equation only ``works" when the train has a particular velocity, that might indeed suggest that the equation does not capture a genuine law, as it would be extremely specific. But, once again, this is not the case. On the contrary, the classical wave equation can be used successfully in many different contexts such as in different locations (e.g., in any lab on Earth or even in a spaceship), in spaces with different velocities (ships, airplanes, trains), and at different times. It is also rather informative, has high predictive value, and supports counterfactuals (e.g., it tells us how some particular wave in a string would have behaved if the string had been pulled in some other way). In short, this equation seems to exhibit all the standard features associated with the equations of natural laws. Furthermore, to say that this equation does not represent a genuine law would open the floor for saying that, for example, the equation for the propagation of electromagnetic waves is also not capturing a law, which would be rather controversial (formally, one can regard the electromagnetic wave equation in vacuum as a three-dimensional version of the classical wave equation).  And, finally, to claim that the classical wave equation is not the equation of a law just seems to be at odds with scientific practice. 

\subsection{Acoustic Waves and the Doppler Effect}
\label{Acoustic Waves and the Doppler Effect}

To end this section, it is illustrative to consider the relationship between acoustic waves and the case of string waves discussed above.  One might wonder, in particular, if the situation involving the vibrations of the string in the train cabin is analogous to cases involving the acoustic Doppler effect. When an object produces waves while it is moving relative to the medium over which the waves are defined, such as in the case of the siren of an ambulance when the ambulance is moving relative to the air, those waves are indeed affected by the relative motion between the source of the waves (the siren) and the medium (the air). In this case, boosts of the ambulance with respect to the air do produce changes in the pitch or frequency of the sound, and so these kinds of boosts do bring about detectable effects. The question, then, is whether this constitutes an example in which EGRP fails analogous to the case of the string waves, or perhaps a novel example in which IGRP also fails.  

Let's start with IGRP. As \citet[\S2]{chengWhyNotSound2021} point out, in the case of classical waves we have to distinguish between the case in which we boost both the object producing the waves (henceforth the ``source") \textit{and} the medium for those waves, and the case in which we boost the object producing the waves without boosting the medium (or the case in which we boost the medium without the source). If one boosts the source without boosting the medium, then a person moving with the source, say the driver of the ambulance, will notice differences in the sound of the siren when the ambulance is at rest compared to when the ambulance is moving. The fact that the person inside the ambulance can detect these differences in the sound of the siren might suggest an analogy to scenarios where a person inside a ship conducts experiments to distinguish between different states of motion within the ship, potentially indicating situations where IGRP may not hold. However, cases in which only the source is boosted relative to the medium seem to be instances in which only a part of the relevant (total) system is being boosted, analogous to boosting just a few objects inside the cabin of Galileo's ship or to boosting the front wheel of a bike without the rest of the bike. Yes, there will be physical effects associated with these kinds of boosts, but IGRP is about cases in which the whole system, not just a part of it, is boosted together with the frame. In other words, when assessing IGRP in the context of classical wave phenomena, we must treat the source and the medium as parts of a total ``wave-system" that must be boosted together as a whole.\footnote{Famously, in the case of light in vacuum there is no medium to be boosted even though many physicists thought that the ether was such a medium (see \citet{chengWhyNotSound2021} for discussion).} And if one does this, the boost does not seem to bring about any physical effects. In the words of \citet[p. 4]{chengWhyNotSound2021}, ``if the medium of propagation is also subject to a Galilean boost, one does not expect violations of the Galilean relativity principle by a given wave" (note that the authors seem to be assuming the internal reading of GRP).  Indeed, Einstein's example of the pipe organ nicely illustrates this point (the organ and the air around it are moving together with the Earth). Thus, acoustic waves do not seem to threaten IGRP.

The previous discussion might suggest that if one treats the source and the medium as parts of a greater system that must be boosted together, then \textit{all} classical waves are compatible with GRP in a Newtonian (non-relativistic) context, even if GRP were to be read along the lines of EGRP. As it turns out, perhaps surprisingly, some classical waves are inconsistent with EGRP even if one boosts the source and the medium together. In fact, the vibrations in a guitar string are a good example! Here the source of the vibration might be a finger plucking the string, and the medium is the string proper.  As we explained earlier in this section,  a camera fixed in the station will see the vibrations as satisfying different laws compared to those seen by the internal camera moving with the system. That is, even when the source (here a person) and the medium (a string) are boosted together relative to the external camera, that camera still detects different laws than those detected by the internal camera (see Figure 1). So even if the source is boosted together with the medium, classical waves can still violate EGRP.\footnote{\label{Classical Wave Lorentz boost} Mathematically, this is captured by the fact that even if the propagation velocity of the wave relative to the medium is constant, as expected when both the medium and source are boosted jointly, Galilean boosts will introduce a time dependence for the wave's position that affects its various derivatives (this time dependence captures how the points along the string move relative to the external camera that is not boosted). This is why the equation in footnote \ref{classicalwave} is \textit{not} invariant under $x \mapsto x-vt$ even if $\sqrt{T/\mu}$, the velocity of propagation of the wave relative to the string, is held constant. Interestingly, the equation is invariant under Lorentz boosts (assuming, again, that $\sqrt{T/\mu}$ is held constant).}

Furthermore, we do not need to focus only on string waves to encounter cases incompatible with EGRP. Certain situations involving acoustic waves also illustrate this point even when we boost the source and the air together. For instance, imagine an ambulance that is parked with the siren on, and consider two detectors, one that is attached to a car moving uniformly relative to the ambulance, and one that is attached to the street near the ambulance. Despite no relative motion between the source (the siren) and the medium (the air) in this scenario, the two detectors will disagree regarding the laws governing the acoustic waves; while the detector at rest will show that the waves obey the standard acoustic wave equation, the detector in motion will indicate that the waves obey a more complex differential equation that takes into account the motion of the detector relative to the medium.\footnote{\label{soundwaveobservermoving} In particular, the equation for acoustic waves in the case of an observer moving relative to the medium with velocity $v$ is 
\begin{equation}
    \partial_t^2 p(\mathbf{x},t)
   + 2 \partial_t\,\mathbf{v}\!\cdot\!\nabla p(\mathbf{x},t)
   + \mathbf{v}\cdot\nabla^2\,\mathbf{x}\,p(\mathbf{x},t)
  = c^2\nabla^2p(\mathbf{x},t),
\end{equation}
where $c$ represents the velocity of the waves with respect to the medium, $p$ the acoustic pressure, $x$ the position and $t$ the time (the standard acoustic wave equation lacks the second and third terms on the left side of this equation).} Hence, as the detectors in the two inertial frames disagree about the law associated with the sound waves produced by the siren, this constitutes a counter-example to EGRP in a non-relativistic context (the detectors also disagree about the frequency, but this is analogous to a disagreement about the dart's speed, and so analogous to a disagreement that is compatible with EGRP). Crucially, in contrast to a case where the ambulance moves relative to the air, in this situation, there are no changes in the velocity between the source and the medium but only changes in the frame from which the source and medium are being studied (from the perspective of the detector in the car, this is a case where the source and the medium are boosted together). 

The upshot is that even though acoustic waves do not threaten IGRP, they do threaten the Galilean Relativity Principle in the form of EGRP, and they do so even when there are no changes in the relative velocity between the source of the wave and the medium (i.e., they threaten EGRP even when the source and the medium are boosted together).

\section{Going Beyond the Classical World}
\label{Beyond the classical world}

Up to this point, I have appealed to some classical systems to show that IGRP and EGRP are different in non-relativistic contexts. In this section, I will generalize the discussion further by considering non-classical systems, both in relativistic and non-relativistic contexts. The upshot, we will see, is that IRP (but not ERP), is an extremely general principle, satisfied by all kinds of systems, including systems that have laws that are not invariant under Galilean or Lorentz boosts (recall from \S\ref{Two senses of having the same laws} that IRP and ERP generalize IGRP and EGRP for the case of all physical systems). This illustrates not only that the phenomenon captured by Galileo's ship is surprisingly universal, but, as we will see at the end of the section (in \S\ref{Galileo's ship and the philosophy of symmetries}), it poses some problems for a widely endorsed interpretation of Galileo's passage in the contemporary philosophical literature on symmetries.

\subsection{Atoms and light}
\label{Atoms and light}

Imagine that in the train cabin, there is an advanced laboratory capable of running the Stern-Gerlach experiment or doing spectral analyses of the hydrogen atom. If we run one of these quantum experiments when the train is parked in the station, and then repeat it when it is moving uniformly at 100 kph in a straight line, we will get the same results, say, the very same distribution of the spectral lines for hydrogen (for now, we are assuming a non-relativistic context). Hence,  we would not be able to distinguish between two uniform states of motion of the train solely based on the outcomes of these quantum experiments.  But if this is so, then it seems that quantum systems satisfy IRP in non-relativistic contexts, just like classical ones do.  
 
Now, it is well known that the Schrödinger equation, used to model non-relativistic quantum systems such as the hydrogen atom at low speeds, is \textit{not} invariant under simple Galilean boosts of the form $x \mapsto x-vt$.\footnote{Rather, it is invariant under more complex boost transformations that require the addition of a term with a phase factor (see \citet{brownGalileanCovarianceQuantum1999} and \citet{greenbergerInadequacyUsualGalilean2001}).} What does this lack of invariance tell us? Something very similar to the case of classical waves! Although this is often left implicit, the Schrödinger equation presupposes a particular frame, namely, a frame in which the so-called ``standing waves" are indeed stationary (these waves are solutions of the time-independent part of the Schrödinger equation). To be more concrete,  when one uses the Schrödinger equation to solve the case of a particle in a box, one assumes a frame in which the box is at rest. It follows that whereas an internal device would see a quantum system in the cabin satisfy the Schrödinger equation, the same would not be true for an external device.\footnote{See \citet{brownAreSharpValues1998} for a discussion of how Galilean boosts of the frame affect the spectrum of a (non-relativistic) quantum system.} Not only is the Schrödinger equation not invariant under Galilean boosts, but as a non-relativistic law, it is also not invariant under Lorentz boosts. So even if the two inertial frames associated with the two states of motion of the train were to be related by a Lorentz boost, the internal and the external camera would still disagree about the laws satisfied by the system in the cabin. It follows that quantum systems modeled by the Schrödinger equation fail to satisfy ERP in both relativistic and non-relativistic contexts.  And yet, just as we could not use the results of a quantum experiment to distinguish between two inertial frames related by a Galilean boost, we also could not do that in the case in which those frames were related by a Lorentz boost. This illustrates that Galilean and Lorentz boosts preserve the outcomes for experiments (confined to the cabin) of systems whose laws are manifestly not invariant under either kind of boost. The phenomenon originally captured by Galileo's ship is remarkably general!

It is also illustrative to briefly consider electromagnetic systems, which are manifestly relativistic. If we perform some experiments involving light inside the train cabin when it is parked, and repeat the experiments when the train is moving uniformly relative to the station, and we further assume that the two frames are related by a Galilean boost, we will obtain the very same results inside the cabin (e.g., the same interference patterns). So Galilean boosts can preserve the behavior of light as seen inside the cabin, exactly in an analogous way to how they preserve the vibrations in a string as seen inside the cabin. But just as it happened with the classical wave equation, the wave equation for light is not invariant under Galilean boosts but under Lorentz boosts. This lack of invariance under Galilean boosts reflects the fact that in a non-relativistic context, a device inside the cabin would disagree with an external device about the kind of behavior of an electromagnetic system inside the cabin---the devices would agree, however, if the two frames were to be related by Lorentz boosts, as Einstein notes in the passage cited in \S\ref{Is IGRP the obvious reading?}. Thus, in non-relativistic contexts, electromagnetic systems can satisfy IRP, but not ERP. In relativistic contexts, these same systems satisfy both IRP and ERP.

In summary,  when assessing if a physical system satisfies the Relativity Principle (or the Galilean Relativity Principle for a classical system), it is important to specify (1) if we mean the internal or the external version of the principle, and (2) if we are assuming a relativistic context in which inertial frames are related to one another by Lorentz boosts, or a non-relativistic context in which they are related by Galilean boosts. For example, in a non-relativistic context, vibrations in a string satisfy IGRP and fail to satisfy EGRP, as the classical wave equation is not Galilean invariant. In a relativistic context, the same vibrations satisfy both IGRP and EGRP, as the classical wave equation is Lorentz invariant.\footnote{This is true if we assume that the velocity of propagation of the wave with respect to the string, $\sqrt{T/\mu}$, is held constant (see footnote \ref{Classical Wave Lorentz boost}).}  Ideal springs satisfy IGRP both in relativistic and non-relativistic contexts but do not satisfy EGRP in either case, as Hooke's law is neither Galilean nor Lorentz invariant.  A quantum system such as a hydrogen atom satisfies IRP in relativistic and non-relativistic contexts, but it does not satisfy ERP in either case (the Schrödinger equation is neither Lorentz nor Galilean invariant). An electromagnetic wave propagating in a vacuum satisfies IRP in relativistic and non-relativistic contexts, but it only satisfies ERP in the relativistic case, as it has Lorentz invariant laws. And so on for other cases. 

The distinction between EGRP and IGRP, on the one hand, and between relativistic and non-relativistic contexts, on the other, highlights a sense in which standard terminology can be rather confusing. For we use ``Galilean" both in the ``Galilean Relativity Principle" and in ``Galilean invariance," as if these terms were always intimately tied to one another and perhaps even equivalent.\footnote{Some authors also use ``Galilean covariance" or ``invariance under Galilean transformations" instead of ``Galilean invariance." But the term ``Galilean transformation" is also ambiguous, as it sometimes refers to Galilean boosts (e.g., see \citet[p. 209]{browneGalileiProposedPrinciple2020}), and sometimes it denotes the more general set of transformations given by (i) spatial and time translations, (ii) spatial rotations and (iii) Galilean boosts (this set of transformations forms the so-called ``Galileo Group"). According to \citet[p. 324]{martinezKinematicsLostOrigins2009}, the first scholar to use the term ``Galilean transformation" was Philipp Frank in 1908 (Frank used it to refer to Galilean boosts in particular).}  But as it should be clear now, the connection between these concepts is rather subtle. In particular, with the risk of repeating myself, note that it is not necessary that a system has laws that are Galilean invariant in order for it to satisfy the internal version of the Galilean Relativity Principle (IGRP), although it is necessary (and sufficient) that it has Galilean invariant laws for it to satisfy the external version (EGRP) in non-relativistic contexts. Furthermore, it is not necessary for a system to have Galilean invariant laws for it to satisfy the external version of the Galilean Relativity Principle in relativistic contexts, as waves in a string illustrate (they have Lorentz invariant laws). In short, we see a hodgepodge of terms appear in the literature with ``Galilean" in the name, making it too easy to lose track of the fact that, as I have argued in this paper, some of these terms actually correspond with very different principles.\footnote{In other words, the fact that two or more principles have ``Galilean" in the name does not imply that they are equivalent, but rather indicates that they are influenced, at least in part, by the work of Galileo. }

\subsection{Galileo's ship and the philosophy of symmetries}
\label{Galileo's ship and the philosophy of symmetries}

To end, let me briefly indicate how the discussion in this section allows us to draw some general lessons that are relevant to recent conversations in the philosophy of symmetries.  In the contemporary literature on symmetries, Galileo's ship has became a paradigmatic example of a case supposedly illustrating how symmetries of the laws (also known as ``dynamical symmetries") give raise to, or are intimately connected with, the preservation of certain observations. To be concrete, authors such as \citet{bradingAreGaugeSymmetry2004}, \citet{dasguptaSymmetryEpistemicNotion2016}, \citet{greavesEmpiricalConsequencesSymmetries2014}, \citet{healeyPerfectSymmetries2009},  \citet{wallaceObservabilityRedundancyModality2022}, \citet{tehGalileoGaugeUnderstanding2015}, and \citet{murgueitioramirezAbandoningGalileoShip2021}---to name a few---all treat Galileo's ship as a paradigmatic case that demonstrates how symmetry transformations of a given system (e.g., Galilean boosts) can both (I) preserve the system's behavior as seen by someone who is transformed together with it, and (II) bring about detectable effects relative to an external system that is \textit{not} transformed (many authors even use the terms ``Galileo-ship-type" or ``Galileo-ship-case" precisely to characterize this kind of situation). For example, think of the case of the dart that was discussed earlier. In that case, a Galilean boost---which is a symmetry of the dart's laws---preserves how the dart's trajectory looks according to the internal camera (which is also boosted), but changes how the trajectory looks with respect to the external camera fixed in the station. What I have shown in this paper, however, is that this characterization of Galileo's ship is just too narrow. There is a vast number of physical systems (not to mention biological systems such as the butterflies Galileo himself mentions!), ranging from classical ones like vibrating strings to quantum ones like atoms, whose laws do not include Galilean boosts as symmetries. And yet, Galilean boosts of these systems still satisfy conditions (I) and (II)---this illustrates that non-symmetry transformations of many systems still satisfy (I) and (I).

This suggests that the phenomenon captured by Galileo's ship has been mischaracterized in contemporary discussions; in its full generality, it is not really about dynamical symmetries, let alone dynamical symmetries preserving certain observations. Rather, it is about the rather general fact that when one boosts a system together with an observer, the  motion of the system relative to the observer is preserved (see \citet[\S7]{murgueitioramirezSymmetriesSprings2024} for a recent proposal along these lines). In fact, this is made clear by Galileo himself! Towards the end of the ship passage, Galileo says that ``the cause of all these correspondences of effects is the fact that the ship's motion is common to all the things contained in it, and to the air also" \citep[p. 186]{galileiDialogueConcerningTwo1967}. What Galileo says here about the objects in the ship can also be said of guitar string vibrations, Einstein's pipe-organ, Stern-Gerlach apparatuses in a lab, and all objects that move together with the Earth as it orbits the Sun. Indeed, the same idea can be applied to \textit{any} system that satisfies IRP (or IGRP) regardless of their symmetries. 

This is not to say that there are no interesting links between observations and dynamical symmetries. In fact, EGRP points to an obvious connection; if Galilean boosts are symmetries of a system, then in a non-relativistic context, an external observer will agree with an observer boosted with the system about the laws (the dart is a good example). But, of course, agreement about the laws is a far cry from agreement about every single observation. In fact, these two observers will disagree about various properties, including the velocity, position, and momentum of the system.\footnote{When a system satisfies EGRP in a non-relativistic context, the internal and external observers will also agree, in addition to the laws, about the value of the properties conserved by the boost. In particular, it follows from Noether's first theorem that, for classical systems, the quantity $x_{CM} - v_{CM} t$ is conserved by Galilean boosts,  where $x_{CM}$ is the center of mass of the system, $v$ the velocity associated with the boost and $t$ the time (for the observer moving with the system, $v=0$) .} On the other hand, if Galilean boosts are not symmetries, the two observers in question will disagree not only about these properties but also about the laws (e.g., the laws for the vibrations in a string). 

In short, the examples discussed in this paper highlight that symmetries do convey interesting information about the external side of things (as captured by EGRP or ERP), but not so much about the internal perspective (as captured by IGRP or IRP). Thus, if one wants to better understand the observations made by someone who is boosted together with the system, as it happens in the ship passage, then less attention should be given to the symmetries of the system, and more to other matters that have been comparatively neglected, such as the relativity and composition of motion (also discussed by Galileo), the rigidity of physical systems (necessary so that they can be boosted without destruction), the relationship between spacetime symmetries---not to be confused with dynamical symmetries---and the behavior of physical systems, and even sociological and historical questions connected to how physicists decide which properties of their theories are observable (e.g., see \citet{readRedundantEpistemicSymmetries2020}).

\section{Conclusion}

The following two assumptions are widely held:

\begin{enumerate}
    \item There is a close relationship between the Galilean Relativity Principle and the symmetries of mechanical systems. 
    \item There is a close relationship between the Galilean Relativity Principle and Galileo's ship thought experiment.
\end{enumerate}

It is very tempting to look at these two assumptions and infer the following: 

\begin{enumerate}
    \item[C.] There is a close relationship between the symmetries of mechanical systems and Galileo's ship thought experiment.
\end{enumerate}

Indeed, the widespread view that there is a strong connection between symmetries and observations seems to be strongly motivated by (C)!  

This paper can be seen as offering an argument that the temptation of inferring (C) from (1) and (2) must be resisted. The argument is simple: (C) is not really motivated by (1) and (2) because the two instances of ``Galilean Relativity Principle"  refer to two different principles, EGRP and IGRP, respectively. The first instance refers to the idea that the laws of mechanical systems are invariant under Galilean boosts, that is, to the idea that boosts are symmetries of these systems. As we explained in \S\ref{Systems that do not satisfy EGRP}, there are exceptions to this principle, such as classical waves and ideal springs. The second instance refers to the idea that if one boosts a system together with an observer (in a straight line), the relative motion between the system and the observer is preserved, and so the observer will not be able to tell apart different states of uniform motion (i.e., different inertial frames) solely based on the system's behavior. This is not primarily about the \textit{laws} obeyed by the system, let alone its symmetries (Galilean or otherwise). In fact, this is not even just about mechanical or classical systems; it works for quantum systems and electrodynamical ones too, even though their laws are not invariant under Galilean boosts!\footnote{Although in quantum cases, perhaps it is more accurate to speak of ``behavior" as opposed to ``motion."} 

In short, the reason that ripples in a pond, a clock pendulum, the sounds produced by the vibrations of a guitar string, experiments at CERN, and virtually the outcome of any experiment performed in a (approximately closed) lab do not depend on the orbital speed of the Earth is \textit{not} that Galilean boosts are symmetries of these systems. Rather, the reason is that the systems, the instruments used to study them, their surroundings, and the scientists themselves are moving together with the Earth. Paraphrasing what Galileo says at the end of the ship passage, the Earth's motion is common to all the things contained within it, including the air.

To end, notice that the principle of the composition and relativity of motion, which allows us to explain why so many different kinds of observations and experiments fail to reveal changes in the Earth's orbital speed, also has the seeds for understanding the even more general idea captured by Newton's Corollary VI (or Einstein's equivalence principle) according to which joint accelerations of a system and an observer preserve how the system seems to behave relative to the observer. And just as it was with the case of simple boosts, we should resist the temptation of inferring that these more general principles ought to be understood via the symmetries of the laws of physical systems. After all, the law of the classical wave equation is \textit{not} preserved under acceleration transformations, yet the waves in the strings of a guitar look the same, as seen by an observer co-moving with the guitar, regardless of whether the spaceship is in free fall or moving uniformly. But a careful discussion of these more general principles will have to wait until another day. 

\section*{Acknowledgements}

I am very grateful to Michael Jacovides, who provided detailed comments on multiple versions of this paper. I also want to thank Jeff Brower, Nicolas Berrío Herrera, Eugene Chua, Martín González Abaúnza, Geoffrey Hall, Jack Himelright, Ben Middleton, and Dana Tulodziecki for their helpful and encouraging feedback. Likewise, I want to thank the audiences for their valuable insights at the three events where I presented an early draft of this paper: the 2024 Pacific APA, the Caltech Philosophy of Physics seminar, and a Philosophy of Physics seminar held at the University of Notre Dame in the Fall of 2023. In addition, I want to thank the referee for their feedback, which helped significantly improve the paper in several ways. Finally, I am immensely indebted to Stacy Sivinski for her editorial work and support. 

\clearpage

\bibliographystyle{apacite}

\end{document}